\documentclass[a4paper]{article}

\usepackage{INTERSPEECH2022}
\usepackage{url}
\usepackage{hyperref}

\usepackage{bm}

\newcommand{\NM}{\mathcal{N}}

\newcommand{\EE}{\mathbb{E}}

\usepackage{xcolor}

\newcommand{\Sec}[1]{section~\ref{sec:#1}}
\newcommand{\Fig}[1]{Fig.~\ref{fig:#1}}
\newcommand{\Table}[1]{Table~\ref{tbl:#1}}
\newcommand{\Eq}[1]{eq.~(\ref{eq:#1})}
\newcommand{\method}[1]{\textbf{#1}}
\newcommand{\mysection}[1]{\vspace{-5pt}\section{#1}\vspace{-3pt}}
\newcommand{\mysubsection}[1]{\vspace{-4pt}\subsection{#1}\vspace{-2pt}}
\newcommand{\mysubsubsection}[1]{\vspace{-2pt}\subsubsection{#1}\vspace{-2pt}}
\newcommand{\revise}[1]{\textcolor{black}{#1}}

\title{End-to-End Text-to-Speech Based on Latent Representation of \\ Speaking Styles Using Spontaneous Dialogue}
\name{Kentaro Mitsui$^1$, Tianyu Zhao$^1$, Kei Sawada$^1$, Yukiya Hono$^2$, Yoshihiko Nankaku$^2$, Keiichi Tokuda$^2$}
\address{
  $^1$rinna Co., Ltd., Japan, 
  $^2$Nagoya Institute of Technology, Japan}
\email{\{kemits,tianyuz,keisawada\}@rinna.co.jp, \{hono,nankaku,tokuda\}@sp.nitech.ac.jp}

\begin{document}

\setlength{\abovedisplayskip}{3pt}
\setlength{\belowdisplayskip}{3pt}
\maketitle

\begin{abstract}
\vspace{-3pt}
The recent text-to-speech (TTS) has achieved quality comparable to that of humans; however, its application in spoken dialogue has not been widely studied.
This study aims to realize a TTS that closely resembles human dialogue.
First, we record and transcribe actual spontaneous dialogues.
Then, the proposed dialogue TTS is trained in two stages:
first stage, variational autoencoder (VAE)-VITS or Gaussian mixture variational autoencoder (GMVAE)-VITS is trained, which introduces an utterance-level latent variable into variational inference with adversarial learning for end-to-end text-to-speech (VITS), a recently proposed end-to-end TTS model.
A style encoder that extracts a latent speaking style representation from speech is trained jointly with TTS.
In the second stage, a style predictor is trained to predict the speaking style to be synthesized from dialogue history.
During inference, by passing the speaking style representation predicted by the style predictor to VAE/GMVAE-VITS, speech can be synthesized in a style appropriate to the context of the dialogue.
Subjective evaluation results demonstrate that the proposed method outperforms the original VITS in terms of dialogue-level naturalness.

\end{abstract}

\mysection{Introduction}

Dialogue is a conversation between two or more people.
In recent years, the development of natural language processing has greatly improved the quality of text-based dialogue generation resulting in human-computer or computer-computer dialogue~\cite{zhang2020dialogpt, thoppilan2022lamda}.
On the other hand, speech is essential for human dialogue.
Therefore, TTS has an important role in facilitating communication between humans and computers.

The development of deep learning has resulted in synthesizing speech in a quality comparable to that of humans~\cite{shen2018tacotron2, kim2021vits}.
However, dialogue speech often has characteristics that are different from those of the recited speech.
First, while recited speech has transcript beforehand, dialogue speech is a spontaneous speech.
Therefore, dialogue speech is more difficult to model than recited speech because of repetition, fillers, prolongation, and breaths.
Second, dialogues are frequently accompanied by backchannels, also known as \textit{aizuchi}~\cite{kita2007nodding} in Japanese, and laughter.
These factors transcribed in the same way can be uttered in various styles.
Thus, it is necessary to appropriately model the one-to-many relationship between text and speech.
Finally, several factors of speech such as pitch~\cite{gravano2014three}, energy~\cite{levitan2011measuring}, and speech rate~\cite{street1984speech} can be in sync with the dialogue partner, which is called entrainment~\cite{levitan2011measuring}.
Considering these features, TTS can resemble more natural human-human dialogue.

Several studies have focused on conversational TTS.
Yokoyama et al.\ used Utsunomiya University spoken dialogue database~\cite{mori2011uudb} to control paralinguistic information~\cite{yokoyama2018effects}.
They utilized paralinguistic information tags and did not consider dialogue history.
Guo et al.\ used the bidirectional encoder representations from
Transformers (BERT)~\cite{devlin2019bert} to compute encodings of current text and chat history and fed them to the encoder of the acoustic model to improve the naturalness of the synthetic speech~\cite{guo2021conversational}.
Cong et al.\ considered the acoustic information of the previous utterance as well as the linguistic information by predicting the Global Style Token~\cite{wang2018gst} of the current utterance from the mel-spectrogram of the previous utterance~\cite{cong2021controllable}.
These two studies used predefined transcript to record spoken dialogue, which may differ from actual dialogue without transcript, in terms of the frequency of the spontaneous behaviors and the presence or absence of backchannels.

In this study, we record a free-form dialogue on a given topic without preparing a transcript to achieve more human-like dialogue speech synthesis.
Of the three aforementioned features of spontaneous dialogue,
(1) we use VITS~\cite{kim2021vits}, an end-to-end TTS which robustly estimates alignment between text and speech by monotonic alignment search (MAS) and blank tokens.
(2) We incorporate an utterance-level latent variable into VITS to facilitate the modeling of one-to-many relationship between text and speech.
Following the framework of VAE~\cite{kingma2014auto}, we propose two methods: VAE-VITS that assumes a standard normal distribution for the prior distribution of the latent variable, and GMVAE-VITS, which assumes a Gaussian Mixture Model (GMM) for the prior.
Furthermore, by sharing the latent space among speakers, training is encouraged to make similar speaking styles between speakers close in the latent space.
(3) We introduce a style predictor that predicts the speaking style of current speech based on dialogue history to realize an entrainment that is close to actual dialogue.
Speech sequences in dialogue history are difficult to handle directly because their length is extremely long.
Therefore, we adopt a two-stage training framework: first, VAE/GMVAE-VITS is trained using a single utterance and then style predictor is trained using a sequence of style representations extracted from past utterances.

\mysection{Spontaneous dialogue corpus}
\label{sec:recording}

To record speech that is close to actual human-human conversation, the following method is used for speech recording and post-processing.
First, two or more speakers are given a topic and asked to talk freely and 
their voices are recorded on independent channels.
Automatic speech recognition (ASR) automatically transcribes the recorded speech.
Transcripts are then manually modified and given time information (start and end time of each utterance) to produce the final transcript with time information.
Using this time information, the audio file is split to obtain utterance-level speech.
Although it is time-consuming to transcribe and assign time information to free dialogue, the use of ASR can greatly reduce the burden of post-processing.
In addition, the lack of predefined transcript allows the speakers to produce more spontaneous speech which contains repetition, fillers, prolongation, and also backchannels.
Speech data recorded in the aforementioned method enables to model the characteristics of actual dialogues more accurately.

\mysection{Two-stage training-based dialogue TTS}

Let $\bm{x}_1, \dots, \bm{x}_N$ and $s_1, \dots, s_N$ be the sequence of dialogue speech and speaker ID of each utterance, respectively.
The purpose of this study is to synthesize the speech $\bm{x}_n$ corresponding to the $n$th text $\bm{t}_n$ and speaker ID $s_n$ by considering the past dialogue speech $\bm{x}_1, \dots, \bm{x}_{n-1}$ and speaker ID $s_1, \dots, s_{n-1}$, that is, to model the distribution $p(\bm{x}_n | \bm{t}_n, s_n, \bm{x}_1, \dots, \bm{x}_{n-1}, s_1, \dots, s_{n-1})$.
In spoken dialogue, each of $\bm{x}_1, \cdots, \bm{x}_{n-1}$ is an extremely long time series, and it is difficult to model them directly when $n$ is large.
Therefore, we propose a two-step training framework which is described in \Sec{method_step1} and \Sec{method_step2}.

\mysubsection{Speaking style modeling using VAE/GMVAE-VITS}
\label{sec:method_step1}
\begin{figure}[t]
\begin{center}
\includegraphics[width=\hsize]{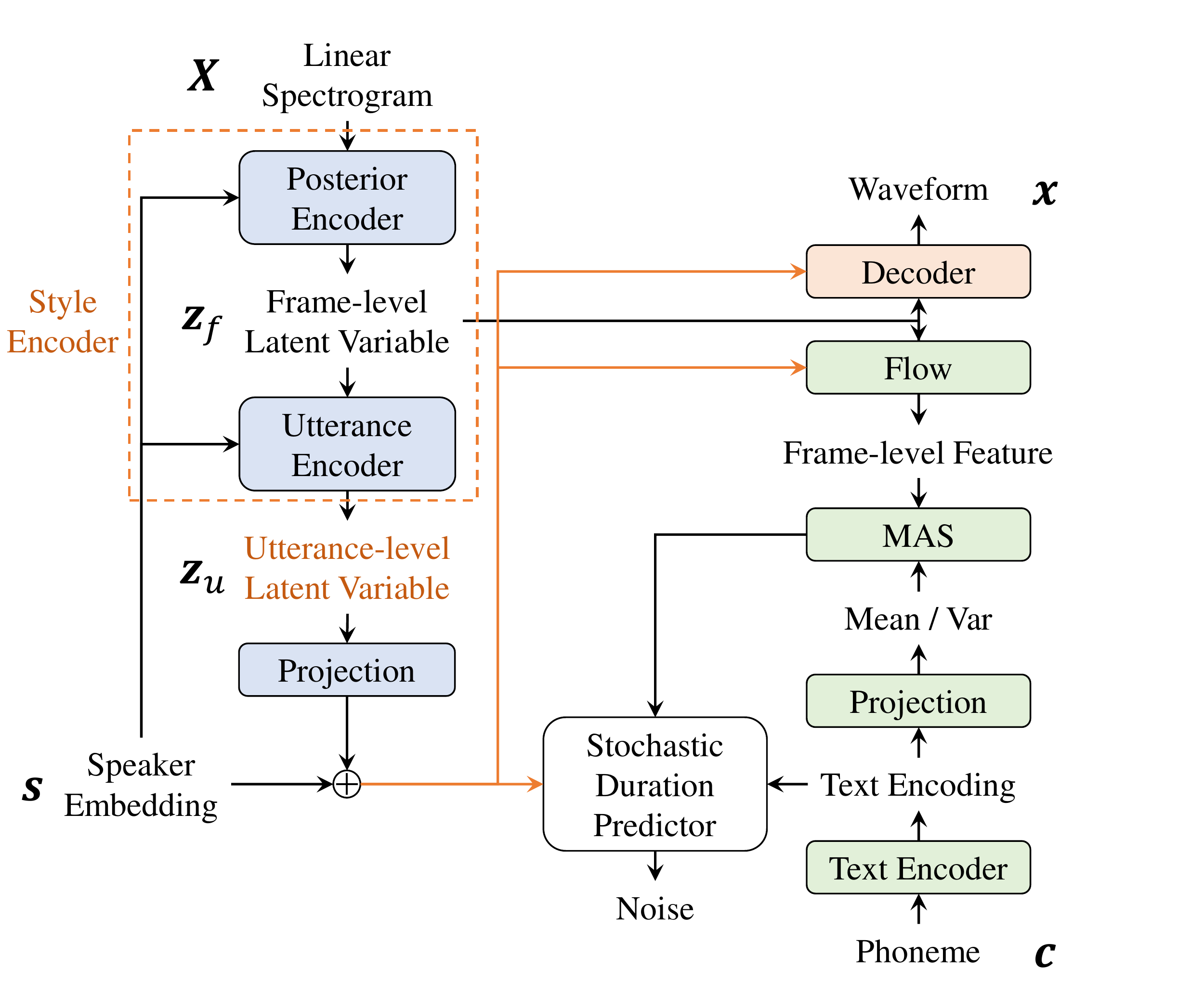}
\vspace{-30pt}
\end{center}
\caption{Training procedure of VITS incorporating utterance-level latent variable.}
\label{fig:vae_vits}
\vspace{-10pt}
\end{figure}

The first training stage models the utterance-level relationship between text and speech, that is, $p_\theta(\bm{x}_n | \bm{t}_n, s_n)$.
In this study, we model this relationship using VITS~\cite{kim2021vits}.
VITS is an end-to-end TTS model that learns the relationship between the phoneme sequence $\bm{c}$ and speech waveform $\bm{x}$ via frame-level latent variable $\bm{z}_f$.
VITS estimates monotonic alignment between $\bm{c}$ and $\bm{z}_f$ during training using 
MAS algorithm~\cite{kim2020glowtts}.
Thus, it can be trained more stably than fully attention-based models such as Tacotron~2~\cite{shen2018tacotron2}.

The proposed method introduces an utterance-level latent variable $\bm{z}_u$ to represent the speaking style of each utterance.
We also introduce an utterance encoder that predicts the mean $\bm{\mu}_u$ and variance $\bm{\sigma}_u^2$ of $\bm{z}_u$ using $\bm{z}_f$ and speaker embedding $\bm{s}$.
That is, the posterior distribution of $\bm{z}_u$ is given as follows:
\begin{align}
    q(\bm{z}_u | \bm{z}_f, \bm{s}) = \NM(\bm{z}_u; \bm{\mu}_u, \mathrm{diag}(\bm{\sigma}_u^2)).
    \label{eq:z_u_pos}
\end{align}
Hereafter, the utterance encoder and the posterior encoder, which predicts $\bm{z}_f$ from the linear spectrogram $\bm{X}$ of speech $\bm{x}$, will be called a style encoder together.
We condition the stochastic duration predictor, flow, and decoder on $\bm{z}_u$ to predict duration, acoustic feature, and waveform considering given speaking style, respectively.
Concretely, we apply a linear projection to $\bm{z}_u$ and add it to the speaker embedding $\bm{s}$, which is fed to each module.
A conceptual diagram of the proposed method is depicted in \Fig{vae_vits}.

The proposed method can be trained by maximizing the evidence lower bound (ELBO) of the following log-likelihood (for simplicity, we omit $\bm{s}$ in the equation below):
\begin{align}
    \log p(\bm{x} | \bm{c})
    &\geq \EE_{q(\bm{z}_f | \bm{x})q(\bm{z}_u | \bm{z}_f)}\left[\log p(\bm{x}|\bm{z}_f, \bm{z}_u)\right] \nonumber \\
    & \quad - \EE_{q(\bm{z}_u | \bm{z}_f)}\left[D_{\mathrm{KL}}(q(\bm{z}_f | \bm{x}) || p(\bm{z}_f|\bm{c}, \bm{z}_u)\right] \nonumber \\
    & \quad - D_{\mathrm{KL}}(q(\bm{z}_u | \bm{z}_f) || p(\bm{z}_u)) \\
    &\approx \frac{1}{M'M''}\sum_{m'=1}^{M'}\sum_{m''=1}^{M''} \log p(\bm{x} | \bm{z}_f^{(m')}, \bm{z}_u^{(m'')}) \nonumber \\
    & \quad - \frac{1}{M''}\sum_{m''=1}^{M''} D_{\mathrm{KL}}(q(\bm{z}_f | \bm{x}) || p(\bm{z}_f|\bm{c}, \bm{z}_u^{(m'')}) \nonumber \\
    & \quad - \frac{1}{M'}\sum_{m'=1}^{M'} D_{\mathrm{KL}}(q(\bm{z}_u | \bm{z}_f^{(m')}) || p(\bm{z}_u))
    \label{eq:vae_vits_elbo}
\end{align}
where $M', M''$ denote the numbers of Monte Carlo sampling for $\bm{z}_f, \bm{z}_u$, respectively.
The first and second terms of \Eq{vae_vits_elbo} can be calculated in the same way as in the original VITS.
Assuming $p(\bm{z}_u)=\NM(\bm{z}_u; \bm{0}, \bm{I})$, the third term is the KL divergence between two multivariate normal distributions, which can be calculated analytically.
We call the proposed method defined by the above model and objective function as VAE-VITS.

This study further examines the use of a Gaussian mixture model (GMM) with equal mixture weights $p(\bm{z}_u|\bm{y}_u) = \NM(\bm{z}_u; \bm{\mu}_{\bm{y}_u}, \mathrm{diag}(\bm{\sigma}_{\bm{y}_u}^2))$ for the prior distribution, following GMVAE-Tacotron~\cite{hsu2019hierarchical}, where
$\bm{y}_u$ denotes the discrete latent class corresponding to $\bm{z}_u$, and the number of latent classes is defined as $K$.
In this case, the ELBO of the log-likelihood is approximated as follows, instead of \Eq{vae_vits_elbo}:
\begin{align}
    \log p(\bm{x} | \bm{c})
    &\geq \frac{1}{M'M''}\sum_{m'=1}^{M'}\sum_{m''=1}^{M''} \log p(\bm{x} | \bm{z}_f^{(m')}, \bm{z}_u^{(m'')}) \nonumber \\
    &\hspace{-12mm} \quad - \frac{1}{M''}\sum_{m''=1}^{M''} D_{\mathrm{KL}}(q(\bm{z}_f | \bm{x}) || p(\bm{z}_f|\bm{c}, \bm{z}_u^{(m'')})) \nonumber \\
    &\hspace{-12mm} \quad - \frac{1}{M'}\sum_{m'=1}^{M'} \biggl\{ D_{\mathrm{KL}}(q(\bm{y}_u | \bm{z}_f^{(m')}) || p(\bm{y}_u))  \nonumber \\
    &\hspace{-12mm} \quad + \sum_{\bm{y}_u=1}^K q(\bm{y}_u | \bm{z}_f^{(m)}) D_{\mathrm{KL}}(q(\bm{z}_u | \bm{z}_f^{(m)}) || p(\bm{z}_u | \bm{y}_u)) \biggr\}.
\end{align}
Some utterances such as backchannels are short and uttered in diverse styles, while others are longer and uttered in a relatively consistent style in dialogue speech.
By assuming GMM for the prior distribution, the various styles of dialogue speech are expected to be represented more accurately.
We call this method GMVAE-VITS.

\mysubsection{Style transition modeling using style predictor}
\label{sec:method_step2}

\begin{figure}[t]
\begin{center}
\includegraphics[width=\hsize]{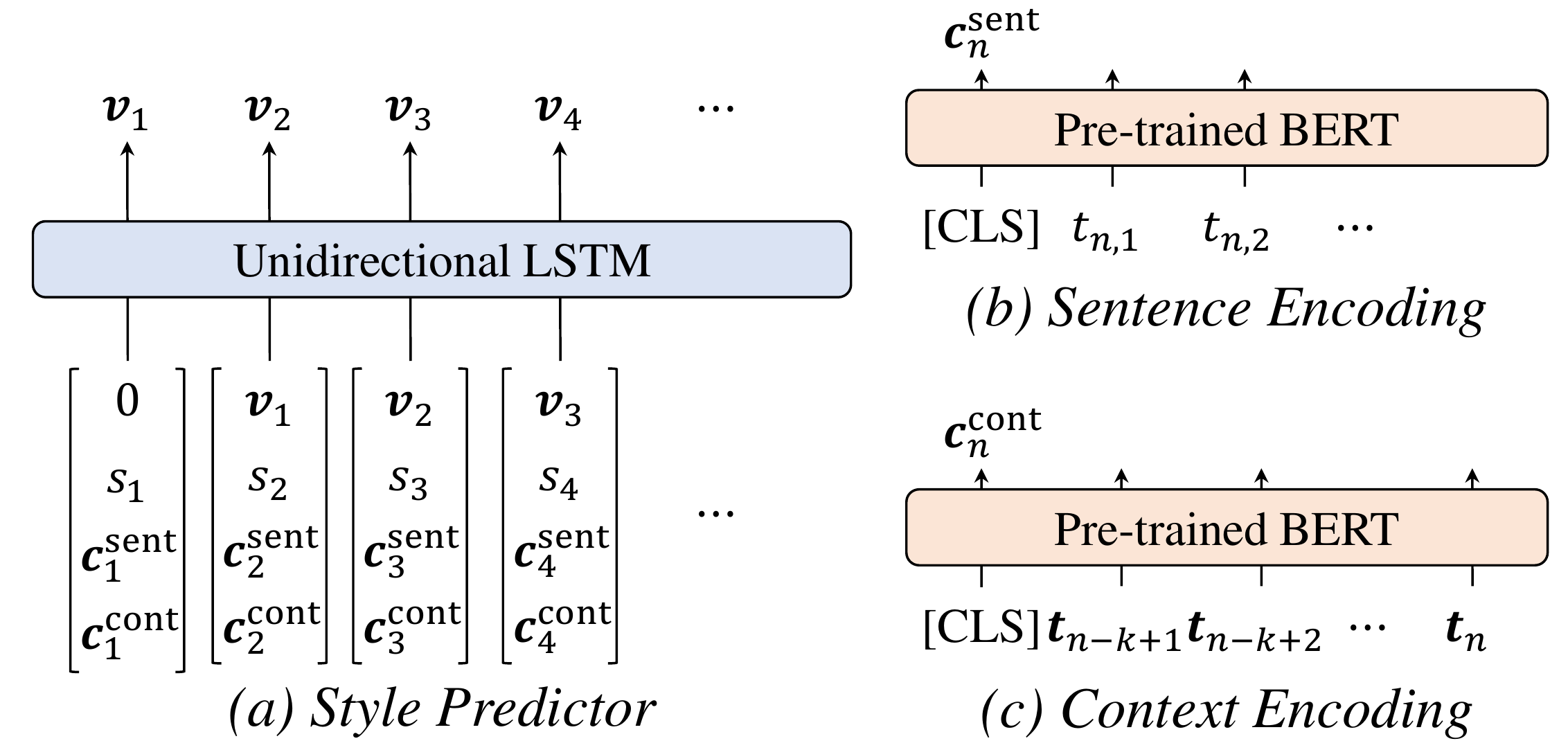}
\vspace{-30pt}
\end{center}
\caption{Conceptual diagram of (a) style predictor, (b) sentence encoding, and (c) context encoding.}
\label{fig:style_predictor}
\vspace{-15pt}
\end{figure}

In the second stage of training, a style predictor which predicts the distribution of $\bm{z}_n$ using sequences of speaking style $\bm{z}_1, \dots, \bm{z}_{n-1}$ and speaker ID $s_1, \dots, s_n$ is trained, with the style encoder trained in the first stage fixed.
An outline of the proposed style predictor is shown in \Fig{style_predictor} (a).
A simple unidirectional LSTM is employed to model the transition of speaking styles during a dialogue.
When we directly use $\bm{z}_u$ described in \Sec{method_step1} as a speaking style representation, the variation caused by sampling from the posterior distribution hinders the training of style predictor.
Therefore, we define $\bm{v}_u = [\bm{\mu}_u^\top, \bm{\sigma}_u^\top]^\top$ as a style vector and used it as an input/output of the style predictor.
Hereafter, we replace the subscript $u$ with the index $n$ in the dialogue for simplicity.

Style predictor takes two sequences as inputs, the style vector sequence $\bm{v}_1, \dots, \bm{v}_{n-1}$ and speaker ID sequence $s_1, \dots, s_n$, and predicts the style vector $\bm{v}_n$ of the current utterance.
The model is trained to minimize the mean squared error $\|\tilde{\bm{v}}_n - \bm{v}_n\|^2$ between the predicted and target style vector $\tilde{\bm{v}}_n$ and $\bm{v}_n$, respectively.
In addition, following Guo et al.~\cite{guo2021conversational}, two types of linguistic features extracted using pre-trained BERT~\cite{devlin2019bert} are input supplementally:
(1) sentence encoding $\bm{c}_n^{\mathrm{sent}}$, an output vector corresponding to the [CLS] token, where the text $\bm{t}_n = (t_{n, 1}, t_{n, 2}, \dots)$ is prefixed with [CLS] token and input to the BERT, and
(2) context encoding $\bm{c}_n^{\mathrm{cont}}$, an output vector corresponding to the [CLS] token, where a series of $k$ text sequences $\bm{t}_{n-k+1}, \dots, \bm{t}_n$ are concatenated, prefixed with a [CLS] token, and input to the BERT.
The procedures of computing these encodings are shown in \Fig{style_predictor} (b) and (c).

\mysection{Experiments}
\label{sec:experiment}

\subsection{Experimental conditions}

\mysubsubsection{Datasets}
Japanese dialogues between two females, who can see each other’s faces through glass and can hear each other's voice with a headphone, were recorded in isolated soundproof chambers by the method described in \Sec{recording} and used for the experiment.
The speakers were colleagues working in radio, which made their conversation more friendly.
The recorded data contained dialogues of 55 topics, which were transcribed and divided into 18,385 utterances.
Azure speech to text was used for ASR.
45 dialogues (15,739 utterances), 5 dialogues (1,284 utterances), and 5 dialogues (1,362 utterances) were used as training, development, and evaluation set, respectively.
All the experiments were conducted using 24~kHz/16~bit speech signals.
A 186-dimensional linguistic feature extracted using Japanese text frontend, Open JTalk\footnote{\revise{\url{http://open-jtalk.sourceforge.net/}}}
, was used as an input of VITS ($\bm{c}$ in \Fig{vae_vits}), instead of phoneme sequences.

\mysubsubsection{Model and training details}

We trained three models: the original VITS~\cite{kim2021vits}, which does not explicitly consider speaking styles, and the proposed VAE-VITS and GMVAE-VITS described in \Sec{method_step1}.
The hyperparameters of VITS were set to be the same as in the previous study.
Following GMVAE-Tacotron~\cite{hsu2019hierarchical}, the utterance encoder was composed of two 1D-convolutional layers with 512 filters and a kernel size of three, two bidirectional LSTM layers with 256 cells at each direction,
and a mean pooling layer followed by a linear projection layer.
The number of Monte Carlo sampling was set to 1 and the dimension of $\bm{z}_u$ was set to 16.
For GMVAE-VITS, the number of latent classes $K$ was set to 10, and the initial value and lower bound of $\bm{\sigma}_{\bm{y}_u}$ were set to $e^{-1}$ and $e^{-2}$, respectively.
All the models were trained for 200k steps using AdamW optimizer~\cite{loshchilov2019adamw} with $\beta_1=0.8, \beta_2=0.99$ and weight decay $\lambda=0.01$.
The batch size was set to 48 and the learning rate was scheduled in the same manner as in the previous study~\cite{kim2021vits}.
KL annealing~\cite{bowman2016generating} was introduced for training VAE/GMVAE-VITS: KL weights of terms newly introduced by the proposed method were increased from 0 to 1 by cosine annealing over the initial 50k steps.

\revise{
Three unidirectional LSTM layers with 256 cells and dropout~\cite{srivastava2014dropout} rate 0.5 were used as the style predictor.
The target style vector was obtained using the style encoder of trained VAE/GMVAE-VITS.
}
We trained BERT~\cite{devlin2019bert} from scratch on approximately 400~GB of Japanese text and used it to compute 1024-dimensional sentence encoding $\bm{c}^{\mathrm{sent}}$ and context encoding $\bm{c}^{\mathrm{cont}}$.
\revise{The text sequence length $k$ for obtaining $\bm{c}^{\mathrm{cont}}$ was set to 10.}
While the dialogues in the training set consisted of 160--693 utterances, we
(1) randomly selected a sequence length $l$ from the range $[10, 30]$, and
(2) randomly extracted consecutive $l$ utterances from each dialogue.
Thereby, we avoided overfitting caused by memorizing the entire series.
The model was trained up to 2,000 steps with batch size 32 using the same AdamW optimizer used in training VITS and the checkpoint with the smallest validation loss was used.

\mysubsection{Results}
\mysubsubsection{Objective evaluation of style predictor}

\begin{figure*}[t]
\begin{center}
\includegraphics[width=\hsize]{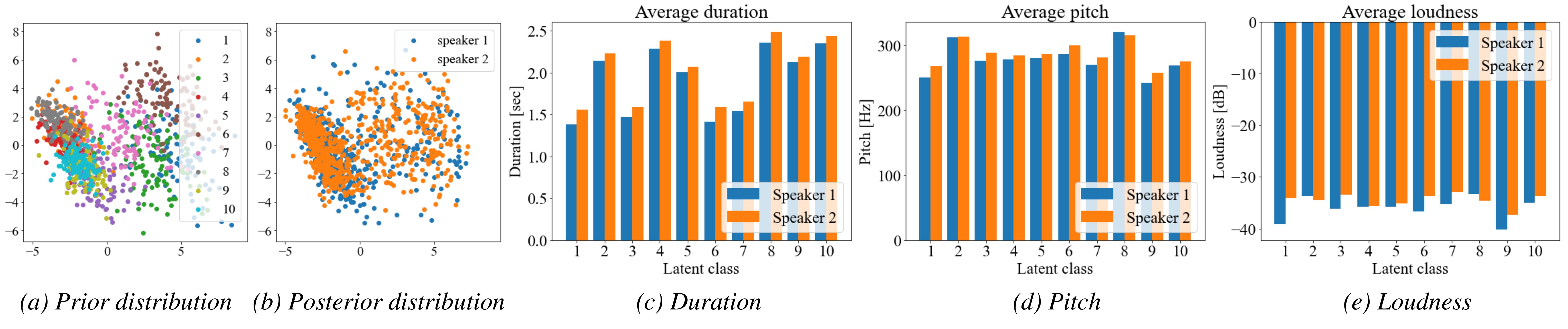}
\vspace{-30pt}
\end{center}
\caption{Analysis of latent space learned by GMVAE-VITS in terms of (a) prior distribution, (b) posterior distribution, (c) average duration, (d) average pitch, and (e) average loudness.}
\label{fig:latent_analysis}
\vspace{-15pt}
\end{figure*}

\begin{table}[t]
\caption{RMSE between the style vector predicted using the style predictor and one extracted from the target speech.}
\renewcommand{\arraystretch}{0.9}
\label{tbl:exp1}
\begin{center}
\vspace{-18pt}   
\begin{tabular}{lcccc}\toprule
Method & \method{None} & \method{S} & \method{C} & \method{S+C}\\\midrule
VAE-VITS & 0.329 & 0.256 & 0.291 & 0.250 \\
GMVAE-VITS & 0.713 & 0.484 & 0.610 & 0.470 \\\bottomrule
\end{tabular}
\vspace{-15pt}   
\end{center}
\end{table}

To evaluate the effectiveness of providing additional linguistic information to the style predictor, we trained the following four models for each of VAE/GMVAE-VITS: 
(1) \method{None}: neither $\bm{c}^{\mathrm{sent}}$ nor $\bm{c}^{\mathrm{cont}}$ was used,
(2) \method{S}: only $\bm{c}^{\mathrm{sent}}$ was used,
(3) \method{C}: only $\bm{c}^{\mathrm{sent}}$ was used, and 
(4) \method{S+C}: both $\bm{c}^{\mathrm{sent}}$ and $\bm{c}^{\mathrm{cont}}$ were used.
The root mean squared error (RMSE) between predicted and target style vectors is presented in \Table{exp1}.
The RMSE of \method{S} was significantly smaller than \method{None} for both VAE and GMVAE, which indicates the effectiveness of $\bm{c}^{\mathrm{sent}}$ in style prediction.
In addition, by comparing \method{None} and \method{C}, or \method{S} and \method{S+C}, we can see that $\bm{c}^{\mathrm{cont}}$ also contributed to improved prediction accuracy.
These results suggest that it is effective to use not only acoustic but also linguistic history in predicting transition of speaking styles during a dialogue.
In the following experiments, we used the style predictor trained in the \method{S+C} condition.

\mysubsubsection{Objective evaluation of overall system}
\begin{table}[t]
\caption{
\revise{Objective evaluation results.}
}
\renewcommand{\arraystretch}{0.9}
\label{tbl:exp2}
\begin{center}
\vspace{-18pt}   
\begin{tabular}{lccc}\toprule
Method & MCD & MSD & DUR \\\midrule
VITS & 7.70 & 9.50 & 0.50 \\
VAE-oracle & 7.06 & 8.18 & 0.44 \\
GMVAE-oracle & 7.04 & 8.17 & 0.41 \\
VAE-predicted & 7.54 & 9.22 & 0.50 \\
GMVAE-predicted & 7.51 & 9.18 & 0.47 \\\bottomrule
\end{tabular}
\vspace{-20pt}   
\end{center}
\end{table}

We conducted an objective evaluation to compare the performance of the proposed methods with baseline VITS.
For the proposed methods, we evaluated two cases: one is to use $\bm{z}_u$ obtained from the target speech using the style encoder (VAE/GMVAE-oracle) and the other is to use $\bm{z}_u$ sampled from the distribution defined by the predicted style vector $\tilde{\bm{v}}_u$ (VAE/GMVAE-predicted).
The performance was evaluated in terms of following metrics:
(1) mel-cepstral distortion (MCD)~\cite{kubichek1993mcd}, the RMSE of a 60-dimensional mel-cepstrum extracted from synthetic and target speech,
(2) mel-spectral distortion (MSD), the RMSE of 80-dimensional mel-spectrogram extracted from synthetic and target speech, and
(3) total duration error (DUR), the error of speech length of synthetic and target speech.
Since the series length differs between synthetic and target speech, dynamic time warping~\cite{berndt1994dtw} was used to align them before calculating MCD and MSD.

The results are presented in \Table{exp2}.
Both VAE/GMVAE-oracle showed significant improvement in MCD and MSD compared to baseline VITS.
DUR was also slightly improved, suggesting that $\bm{z}_u$ represents duration-related features as well as acoustic features.
VAE/GMVAE-predicted also showed improvement in MCD and MSD relative to baseline VITS.
This indicates that the style predictor was able to predict speaking styles that are close to those of target speech.
The performance of GMVAE-VITS was slightly better than VAE-VITS for both oracle and predicted.
This is probably because the richer prior of GMVAE-VITS could represent the various speaking styles of dialogue speech more appropriately.
In the next section, we further compare the proposed GMVAE-VITS with baseline VITS.

\mysubsubsection{Subjective evaluation of overall system}
\begin{table}[t]
\caption{Results of MOS evaluation on utterance-level and dialogue-level naturalness with 95\% confidence intervals.}
\renewcommand{\arraystretch}{0.9}
\label{tbl:exp3}
\begin{center}
\vspace{-15pt}   
\begin{tabular}{lcc}\toprule
Method & Utterance & Dialogue \\\midrule
VITS & 3.38$\pm$0.14 & 3.34$\pm$0.12  \\
GMVAE-oracle & 3.51$\pm$0.12 & 3.59$\pm$0.12 \\
GMVAE-predicted & 3.56$\pm$0.12 & 3.53$\pm$0.11 \\\bottomrule
\end{tabular}
\vspace{-20pt}   
\end{center}
\end{table}

We conducted two mean opinion score (MOS) tests to evaluate the subjective quality of the synthetic speech\footnote{Speech samples are available at : \url{https://rinnakk.github.io/research/publications/DialogueTTS}.}.
In the utterance-level evaluation, raters were presented only one utterance and asked to evaluate its naturalness.
In the dialogue-level evaluation, raters were presented a short dialogue consisting of 6 utterances (approximately 10--20 sec) and asked to evaluate its naturalness as a spoken dialogue (whether natural entrainment occurred, whether the speaking style was suitable for the context, etc.).
\revise{
We used ground truth timing of each utterance to construct dialogue samples because our spontaneous dialogue corpus contained numerous overlaps and simply playing the synthesized speech alternatively resulted in unnatural dialogue.
For dialogue-level evaluation, we also computed text-speech alignment over recorded speech using MAS and used the alignment information for synthesis to align speech length with the original one.
}
The evaluation was conducted on a 5-point scale from 1 (bad) to 5 (excellent). 
Thirty raters participated in the evaluation, and each rater evaluated thirty speech samples.

The results are presented in \Table{exp3}.
With regard to utterance-level naturalness, although the scores of GMVAE-oracle/predicted were slightly higher than VITS, there was no significant difference between them.
Though VITS does not utilize explicit style representation, the synthetic speech was evaluated as natural because various speaking styles exist that sound natural when heard as a single utterance.
Regarding dialogue-level naturalness, the score of GMVAE-oracle was significantly higher than VITS ($p=0.003$ \revise{in Student's t-test}), confirming that using appropriate speech styles contributed to the naturalness of dialogue.
Furthermore, GMVAE-predicted also achieved a significantly higher score than VITS ($p=0.021$), indicating that style predictor was able to predict the appropriate speaking style when heard as a dialogue.

\vspace{-2pt}
\mysubsubsection{Analysis of latent space}
\Fig{latent_analysis} (a) and (b) illustrates the prior and posterior distribution of trained GMVAE-VITS, respectively, where dimensionality reduction was applied using principal component analysis.
We observed that the latent representations of the two speakers were mixed, indicating that the learned latent space was speaker-independent.
This is because the speaker embedding $\bm{s}$ was explicitly used, allowing the latent variable $\bm{z}_u$ to represent only speaker-independent speaking styles.
We also synthesized all texts in the evaluation set with different speaker IDs and latent classes and described the average duration, pitch of voiced segments, and loudness of synthetic speech in \Fig{latent_analysis} (c), (d), and (e).
We confirmed that each latent class had different characteristics and they were common across speakers.
With these characteristics, the learned prior distribution can be applied to modify speaking style to the desired one.

\vspace{-2pt}
\mysection{Conclusions}

In this study, we aimed to synthesize spoken dialogue that is close to human spontaneous dialogue and proposed
(1) recording and transcription of free-form dialogues without transcripts,
(2) VAE/GMVAE-VITS to model various speaking styles, and
(3) a style predictor that predicts speaking styles using linguistic and acoustic features from past dialogues.
The combination of GMVAE-VITS and the style predictor achieved higher naturalness than conventional VITS in a dialogue-level evaluation.
The latent space acquired by GMVAE-VITS was speaker-independent and had different characteristics for each latent class.
This study assumed that transcriptions of past utterances and timing of each utterance were available; however, actual applications will require estimating these as well.
Future work will include introducing a mechanism to automatically estimate them and unifying the proposed two-stage training framework into a single end-to-end training framework.

\bibliographystyle{IEEEtran}

\bibliography{ref/ml,ref/nlp,ref/speech}

\end{document}